\documentclass[a4paper,fleqn]{cas-dc}

\usepackage[numbers]{natbib}
\usepackage{amssymb}
\usepackage{multirow} 
\usepackage{graphicx}
\usepackage{fontawesome5}
\usepackage{rotating} 
\usepackage{caption}
\usepackage{array}
\newcolumntype{M}[1]{>{\raggedright}m{#1}}
\usepackage{booktabs}
\usepackage{makecell}
\usepackage{longtable}
\usepackage{lscape}
\usepackage{tikz}
\usepackage{longtable}
\usepackage{hyperref}
\usepackage{pdflscape} 
\usepackage{lscape}
\usepackage{rotating}
\usepackage{makecell}
\usepackage{bbding}
\usepackage{multirow}
\usepackage{geometry}
\usepackage{afterpage}
\usepackage[utf8]{inputenc}
\usepackage{supertabular,booktabs}
\usepackage{ragged2e}
\usepackage{natbib}
\renewcommand\harvardurl[1]{\textbf{URL:} \url{#1}}
\usepackage{hyperref}
\usepackage{url}

\usepackage{breakurl}
\usepackage{tabularx}
\usepackage[super]{nth}
\usepackage{algorithm}%
\usepackage{algorithmicx}%
\usepackage{algpseudocode}%
\usepackage{listings}%
\usepackage{subcaption}
\usepackage{xcolor}
\usepackage{comment}
\usepackage{adjustbox}
\usepackage{mathrsfs}
\DeclareMathAlphabet{\mathscr}{OMS}{rsfs}{m}{n}

\begin{document}
\shorttitle{A Hybrid Framework For Crypto-Ransomware Detection In Enterprise Shared Storage}    


\title [mode = title]{A Hybrid Framework For Crypto-Ransomware Detection In Enterprise Shared Storage}  

\tnotemark[1] 


%

\author[1]{Gervais Hatungimana}
\cormark[1]
\ead{G.Hatungimana@latrobe.edu.au}
\author[2]{Abdun Naser Mahmood}

\author[3]{Mohammad Jabed Morshed Chowdhury}

\cortext[1]{Corresponding author}

\affiliation[1]{
    organization={Department of Computer Science and Information Technology, La Trobe University},
    city={Melbourne},
    state={VIC},
    country={Australia}
}

\begin{abstract}
\abstract{Most corporate workplace environments enforce policies and technical controls that limit the storage of sensitive data on client endpoints. Consequently, ransomware operators have evolved variants that expand their attack surface from local systems to network drives and shared storage resources. As traditional endpoint detection mechanisms focus primarily on local system behaviour, a compromised client can impact remote file servers, such as by encrypting shared data, without directly triggering behavioural changes on the servers themselves. In this paper, we propose a hybrid detection framework for detecting crypto-ransomware intrusion within integrated file server and client environments. The framework is based on a new technique referred to as  Region of Interest (RoI) to analyse network traffic and extract Indicators of Compromise (IoCs). The IoC repository serves as an additional ruleset to enhance existing security tools such as EDRs and IDSs, while RoI-derived features are used to train an ML model to detect highly evasive variants. This study incorporates a broader set of ransomware families and carefully selected benign behaviours based on domain expertise, ensuring coverage of common user actions that could interfere with ransomware detection. Beyond IoCs, which operate in a signature-based manner, our machine learning module achieves a detection precision of 99.64\%, with a 0\% false negative rate (FNR) and a minimal false positive rate (FPR). Furthermore, the proposed method enables early detection, identifying ransomware intrusions before significant damage occurs, achieving an accuracy of 99.44\%.}
\end{abstract}
\begin{keywords}
 Ransomware detection\sep network traffic analysis\sep Indicators of Compromise\sep machine learning \sep early detection \sep evasive ransomware \sep endpoint protection\sep intrusion detection\sep cybersecurity
\end{keywords}

\maketitle

\section{Introduction}

Although ransomware is often considered a recent phenomenon in cybersecurity, the concept of taking computer data hostage dates back to 1989 with the creation of the AIDS Trojan. Developed by Dr Joseph Popp, an evolutionary biologist and Harvard University alumnus, the Trojan was distributed following the World Health Organisation (WHO) AIDS conference in Stockholm. Popp sent infected floppy disks by mail to conference attendees, marking one of the first known instances of malicious software designed to extort individuals for financial gain.
Victims of the AIDS Trojan were initially asked to pay approximately \$189 to regain access to their encrypted data \cite{smith1990aids}. Since then, new variants and ransomware families have emerged each year, evolving in both complexity and impact. Unlike traditional computer viruses, ransomware typically does not cause permanent damage to the operating system. However, it can disable certain system services, often rendering it unusable until the victim pays the ransom. One of the most prevalent types of ransomware is crypto-ransomware, which encrypts the victim’s data and demands payment in exchange for the decryption key. From a destructive standpoint, crypto-ransomware remains one of the most significant threats due to its ability to lock critical data.

The strategies to combat ransomware attacks involve a combination of electronic and non-electronic methods. Non-electronic strategies primarily focus on educating users about cybersecurity threats, with a particular emphasis on social engineering techniques used by cybercriminals. Electronic strategies, on the other hand, involve the deployment of specialized software on target endpoints to prevent the successful execution of ransomware or other forms of malware.

So far, electronic methods have proven to be the most reliable defence against ransomware attacks. However, the software tools currently used to detect ransomware often fall short of expectations, particularly due to the novel and advanced techniques employed by cybercriminals, which lead to Zero-Day attacks. To mitigate the challenges posed by zero-day vulnerabilities, numerous scholarly articles, such as those presented in \ref{tab:1}, suggest leveraging the power of Machine Learning (ML) to enhance the effectiveness of existing detection tools.

In addition to Zero-Day attacks, ransomware operators have increasingly adopted advanced techniques and hacking practices to compromise their targets and ensure the successful execution of ransomware payloads. This practice is commonly referred to as "evasion.". To counter these evolving threats, it is recommended that detection tools be enhanced to identify incidents without relying solely on the host system's state. One potential approach is to expand the scope of detection by incorporating other sources of information, such as network traffic, which can provide additional insight into the attack.

Most corporate workplace environments enforce policies and technical controls that limit the storage of sensitive data on client endpoints. One of the widely used techniques used for file sharing in the Windows environment is the Server Message Block (SMB) protocol. 

SMB was developed in 1983 to enable file sharing between computers on a network, allowing remote devices (servers) to make files accessible to clients over a TCP/IP connection. Since its inception, SMB has undergone several improvements. SMBv1 was implemented with Windows Server 2003 but was found to be vulnerable to the EternalBlue exploit and was subsequently deprecated in 2003. SMBv2 replaced SMBv1 in Windows Server 2008 and Windows Vista, and is still widely used despite being susceptible to eavesdropping vulnerabilities because as all messages exchanged between the server and the client are transmitted in plain text. 

The latest version, SMBv3, supports data encryption to address these security concerns. Both SMBv2 and SMBv3 are still supported and actively used. \cite{SMB1vsSMB2}. For instance, contemporary Microsoft operating systems, including Windows 10, Windows 11, and Windows Server versions such as 2019, 2022, and 2025, enable SMBv2 and later versions by default and negotiate the highest mutually supported SMB dialect during communication. 

In addition to general-purpose IT systems, SMBv2 is commonly encountered in legacy and constrained environments, including Internet of Things (IoT) and Operational Technology (OT) systems, where long device lifecycles, vendor dependencies, and limited firmware update capabilities often delay adoption of newer SMB dialects. SMBv2 is also widely implemented in non-Windows platforms, such as Samba-based systems and network-attached storage devices, enabling interoperability across heterogeneous networks. 

\begin{figure}
    \centering
    \includegraphics[width=0.75\linewidth]{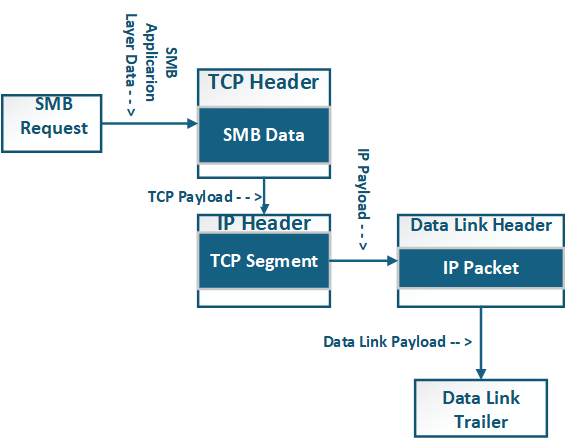}
    \caption{SMB Encapsulation }
    \label{fig:1}
\end{figure}

SMB data consists of commands initiated by the client to access and manipulate remote files hosted on a remote server. The SMB data structure includes various headers that carry different commands \cite{opcode}. As data move through the layers of the OSI model \cite{Encapsulation}, they undergo an encapsulation process. As depicted in Fig.\ref{fig:1}, the SMB data is passed to the transport layer, where it is divided into smaller chunks, known as segments or streams, based on the maximum number of bytes that can be transmitted at once. The size of these segments is determined through a negotiation process governed by TCP standards. At the transport layer, the TCP header is added to each segment, which is then passed to the network layer. The network layer adds IP addresses to the segment, ensuring that they can be routed across the network. At this point, the segment is called a packet, which is forwarded to the data link layer for final delivery.

A typical use case for this protocol is an environment consisting of a File Server and multiple Client PCs. Client PCs access files hosted on the File Server via mapped network drives.

In this paper, we propose a hybrid detection framework for detecting crypto-ransomware intrusion within enterprise Shared Storage. This approach improves detection accuracy and robustness and enables timely alerts in the earliest stages of the attack, thus enhancing the overall response to ransomware threats. 

While research on ransomware detection in file‑share environments is not new, previous studies still have several limitations, such as:
\begin{itemize}
\item \textbf{Risk of dataset and environmental bias:} 
The inclusion of real-user traffic as a benign baseline does not guarantee environment neutrality. User behaviour is strongly influenced by organisational context, system usage patterns, and access controls, which may limit the comparability of benign and malicious traffic across different environments. Furthermore, there is no assurance that legitimate user activity within the observed environment is capable of generating behavioural patterns comparable to ransomware operations, particularly with respect to file access rates, directory traversal, and network share interactions. Consequently, the detection model may capture environment-specific workload characteristics rather than invariant indicators of ransomware behaviour.

\item \textbf{Limited reproducibility and comparability:} 
Network traffic collected from real production environments is typically subject to privacy and confidentiality constraints and is therefore unlikely to be shared. The unavailability of such datasets limits the ability of other researchers to reproduce the results, independently evaluate the proposed technique, or perform fair comparisons with alternative approaches.

\item \textbf{Partial Dataset:} The lack of datasets containing both malicious and benign samples remains a significant challenge for ransomware detection research.

\end{itemize}

A comparison of existing scholarly articles and our approach is summarised in Tab.\ref{tab:1}. The limitations identified in previous research and the reasons why this study is necessary can generally be grouped as follows:

\subsection{Contributions}
In this sub-section, we highlight our contributions towards addressing issues as previously discussed.
\begin{table*}[b]
\centering
\caption{Comparison of our design and closely related works}
\label{tab:1}
\setlength{\tabcolsep}{4pt}

\begin{tabular*}{\textwidth}{@{\extracolsep\fill}lcccccc}
\toprule
\textbf{Work} 
& \multicolumn{3}{c}{\textbf{DESIGN}} 
& \textbf{Environment Neutral}
& \textbf{Reproducibility}
& \textbf{Full Dataset} \\
\cmidrule(lr){2-4}
\cmidrule(lr){5-7}

& \begin{turn}{90}\makecell{Technique}\end{turn}
& \begin{turn}{90}\makecell{Features \\ Engineering \\ Techniques}\end{turn}
& \begin{turn}{90}\makecell{IoC-based \\ Detection}\end{turn}
& \begin{turn}{90}\makecell{}\end{turn}
& \begin{turn}{90}\makecell{}\end{turn}
& \begin{turn}{90}\makecell{}\end{turn} \\
\midrule

{\cite{Berrueta2019}}       
& Threshold 
& 1 Sec. Window
& \XSolid 
& \XSolid 
& \XSolid
& \XSolid \\

{\cite{EduardoBerrueta2022}} 
& ML 
& 1 Sec. Window 
& \XSolid 
& \XSolid 
& \XSolid
& \XSolid \\

\textbf{Ours}              
& ML 
& RoI 
& \Checkmark
& \Checkmark 
& \Checkmark
& \Checkmark \\

\bottomrule
\end{tabular*}
\end{table*}

\begin{enumerate}
    \item 
A new feature engineering technique referred to as the Region of Interest (RoI). While numerous scholarly articles rely on flow-based approaches \cite{Netflow} and time window methodologies, each of these approaches has its inherent limitations, as outlined earlier in this section. The RoI technique mitigates these limitations by using clear demarcation codes to define specific data samples, thus reducing the potential for bias in design. This method is both simple and reproducible, enabling more effective comparisons of results across various studies.
\item
A new hybrid framework combining machine learning (ML) techniques with Indicators of Compromise (IoCs) to detect known and unknown ransomware at an early stage. The framework integrates newly discovered network-based IoCs with network perimeter devices to enhance ransomware detection capabilities. Experimental results demonstrate that the machine learning model is capable of identifying unknown ransomware with high accuracy and a low false negative rate. The synergy between these components provides a robust tool for mitigating ransomware attacks, offering significant potential to improve cybersecurity defences.

\item
We present two new datasets to advance ransomware detection. The first supports machine learning, covering 22 ransomware families and 15 benign apps selected for behavioural similarity to ransomware to reduce bias. The second contains ransomware  Indicators of Compromise (IoCs) from these families. These datasets offer value to both research and industry.
\end{enumerate}
\section{Related Works}

In this section, we review various techniques proposed in previous scholarly articles for ransomware detection in a file-server environment. 

The REDFISH algorithm was proposed in \cite{Berrueta2019}. Morato et al. analysed the SMBv2 protocol in network traffic to design a ransomware detection algorithm in file-sharing environments. The study involved 50 samples from 19 ransomware families. The algorithm detects ransomware activity early by analysing file-sharing network traffic, specifically SMB operations on shared directories. It monitors the sequence and timing of file read, write, and deletion operations, looking for patterns characteristic of ransomware, such as rapid reading of files followed by immediate writing of encrypted content. REDFISH raises an alarm when a predefined number of file deletions ($N$) occur within a short time window ($T$) and the average read/write throughput exceeds a threshold ($V_{\text{thres}}$). By correlating these events in real time, the algorithm identifies ransomware activity before significant data loss occurs, without requiring any software on the endpoints. A significant limitation of the REDFISH algorithm is that it requires parsing all network traffic to extract file paths and SMB commands, which can introduce substantial computational overhead, particularly in high-throughput networks. Furthermore, the algorithm assumes certain file and directory structures in the shared environment; variations in organisational layouts or folder hierarchies may impact the accuracy of early detection. Additionally, the method relies on visibility of file-sharing traffic, so encrypted or obfuscated SMB sessions could hinder its effectiveness.

A Machine Learning (ML) model was recently proposed in \cite{EduardoBerrueta2022}. E.Berrueta et al. (2022) extended SMB traffic with a focus on the newer SMBv3 protocol. The study correlates TCP packets exchanged between the client and server with ransomware-related file operations, including read, write, and other command activities. These events are aggregated within one-second time windows, resulting in dataset features that represent the accumulated byte volumes across different time intervals. While different ML model been tested with different time intervals, Neural Networks (NE) maintains the accuracy over 99\%.

\section{Proposed Method }
Based on gaps towards ransomware detection as discussed previously, this section presents details about the Hybrid framework for detecting crypto ransomware in a file share environment. 

The framework heavily relies on packet codes referred to as packets of interest ($PoI$) and Indicators of compromise ($IoC$). We extracted codes for encrypted (SMBv3) and non-encrypted (SMBv2) environments. However, in order to maintain paper conciseness, we present the framework using SMBv2 codes. We keep equivalent codes for SMBv3 in a separate IoC dataset. The main modules of this framework include  Per Packet IoC Detection, Per RoI Detection and ML-Based detection.

\begin{figure*}
    \centering
    \includegraphics[width=1\linewidth]{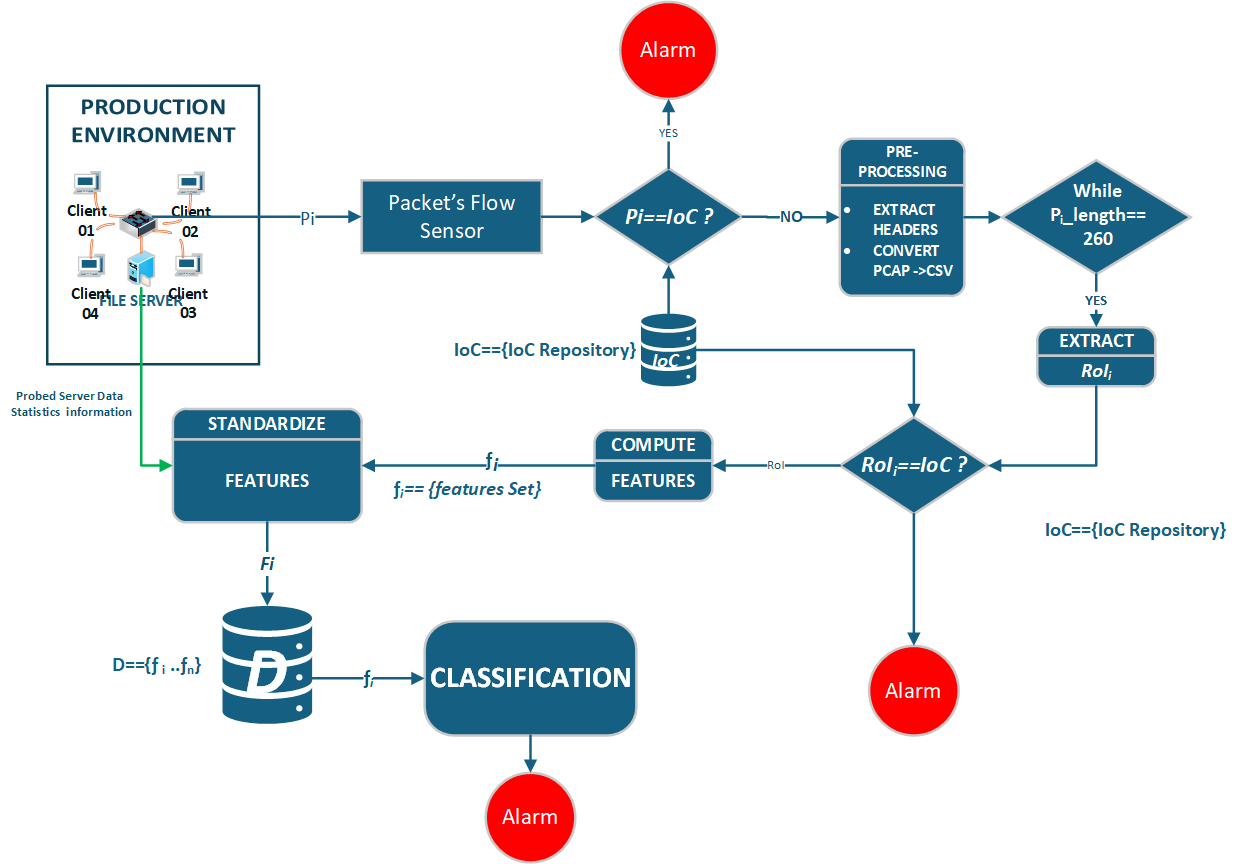}
    \caption{Hybrid Detection Framework}
    \label{fig:2}
\end{figure*}

As depicted in Fig.\ref{fig:2}, the combination of these components enables early detection, automatic remediation, and effective incident management. The system inspects each network packet for well-known behaviours using IoC. If the packet matches any IoC, indicating a known attack, the system immediately triggers automatic remediation, halting further processing. If the packet does not contain any IoC, it is then passed through the Region of Interest (RoI) process. This method analyses the packet's context and behaviour to identify suspicious activity or unknown attack patterns. If RoI analysis identifies the packet as malicious, the system alerts the Security Operations Centre (SOC) team for further investigation and response.

\subsection{Per Packet IoC Based Detection Phase}

This module scans individual packets against known indicators, such as the presence of extortion message drops, and initiates immediate remediation upon detecting a match. The ransomware must inform the victim about the damage and provide instructions on how to make payments(\textbf{extortion}). The commonly used way to do that is by dropping extortion messages in one or more folders on the victim's device. It is the only file that does not get encrypted. The length or size of such a file depends on the information it contains; it is different from family to family. Since each family has a unique message size, it becomes an Indicator of Compromise (IoC) for well-known ransomware families. In experimentation, the message size can be manually depicted through the file properties and then stored in the $IoC$ repository for reference during live traffic inspection. Some characteristics of such files we found so far include:
\begin{itemize}
    \item It appears once in one folder
    \item It is always a result of $C410$, a newly created file
    \item It is the same size across all folders
    \item It can be split into  small chunks to evade detection
\end{itemize}
We use the procedure in algorithm \ref{alg:1} to inspect the network packet. For each network packet subsequent to the SMBv2 packet of length 410, if the packet length value matches any $IoC$ that we collected, that packet is malicious, and an alarm or corrective action is triggered. If nothing is matched, that packet proceeds to the next phase.
\begin{algorithm}
\caption{Per Packet $IoC$ Based Detection Procedure }
\label{alg:1}
\begin{algorithmic}[1]
\State Let $ P_i =   Forward \quad packet$
\State Let $ l = P_i \quad length \quad header$
\For{each $P_i[l]$}   
    \State \hspace{1em} $\ If (\theta P_i[l] \in IoC)$ 
        \State \hspace{2em} $P_i==Malicious$
         \State  end if
    \EndFor
\end{algorithmic}
\end{algorithm}

\subsection{Per RoI, IoC Based Detection Phase}
Certain variants of evasive ransomware attempt to conceal their extortion messages by fragmenting them into multiple packets. When individual packets fail to match indicators of compromise (IoCs) in earlier modules, the system employs Region of Interest (RoI) techniques to identify potential ransomware extortion message drops that may have been fragmented as an evasion tactic. The details about extracting $RoI$ are explained in section \ref{RoI Technique}.

As depicted in algorithm \ref{alg:2}, for each unique value of the Length feature in $RoI$ forward packets, if the ratio between the sum of all packets with the same length value and itself matches any $IoC$, then $RoI$ is malicious and alarm or correction action is triggered. If nothing is matched, the $RoI$ proceeds to the last checkpoint via Machine Learning.

\begin{algorithm}
\caption{Per $RoI$  $IoC$ Based Detection Procedure }\label{alg:2}
\begin{algorithmic}[1]
\Require $ FoI \subset RoI$; A Set of forward  packets flow after $rp410$ packet 
\State Let $RoI\left[l\right]  = \{l_i \}, where \quad i\in \{1,...,n\}$; $n == \forall RoI$ and $l$ == Length  feature value 
\For{each $l_i\in RoI$}     
\State \hspace{1em} $\ If (\frac {(\sum \theta l_i )} {\theta l_i} \in IoC$) Then
    \State \hspace{2em} $RoI=Malicious$
     \State  end if
\EndFor
 \end{algorithmic}  
\end{algorithm}

 \subsection{ML-Based Detection Phase}   

This module is used to detect novel ransomware variants by analysing Region of Interest (RoI) packets for anomalous behaviours that deviate from established patterns. These behaviours may include the presence of encrypted data or irregular access to shared files, which are indicative of potential ransomware activity.
 
Machine learning (ML) plays a pivotal role in artificial intelligence (AI) by enabling computers to mimic human thinking. ML involves various algorithms that classify data using techniques such as strata, probability, or decision trees. Each ML technique employs multiple methods to capture and manipulate patterns within existing data, allowing for the classification of unseen data. Well-known examples of these methods include Support Vector Machines (SVM), Neural Networks, Logistic Regression, Discriminant Analysis, Random Forests, Linear Regression, Naïve Bayes, K-nearest neighbours, and Decision Trees. Notably, Neural Networks have evolved into an independent field known as Deep Learning, which has garnered significant attention in AI research. The choice of which ML technique or method to use depends on the nature and complexity of the data being analysed. 

A common characteristic across all ML methods is their reliance on a benchmark dataset, which serves as the basis for training the algorithms. Once a benchmark dataset is established, the next critical step is feature selection, which involves determining which data patterns should be used to train the model. This step marks a key distinction between traditional machine learning and deep learning, particularly when dealing with large datasets. In traditional ML, human intervention is often required to select the most relevant features for training. In contrast, deep learning models, particularly neural networks, can autonomously identify which features are most relevant, minimising the need for manual feature selection. In this framework, we use \textbf{Random Committee} Machine Learning Model. The decision of this Random Committee over other ML techniques is after experimentation results as discussed in section 6.

\section{Experimentation} \label{experimentation}

In this section, we present the techniques used for carrying out the experimentation, including sample sources, execution, capturing network traces, and processing. 

\subsection{Dataset}
E.Berrueta \cite{Berrueta2020OpenRepository} developed and published a public data repository containing the results of executing over 70 ransomware samples collected between 2015-2020 from different families.  The repository includes network traffic captures (DNS and TCP) and file Input/Output (I/O) operations generated by each sample during its activity, including the read and written bytes, the time between open operations, the number of deletions, and the file sizes. The repository comprises ransomware families collected between 2015 and 2020. While valuable for benchmarking and reproducibility, it may not capture the behaviour of more recent ransomware strains and evolving attack techniques. Therefore, the development of updated datasets remains necessary to support contemporary studies and the evaluation of modern detection approaches.

\begin{figure*}
    \centering
    \includegraphics[width=0.75\linewidth]{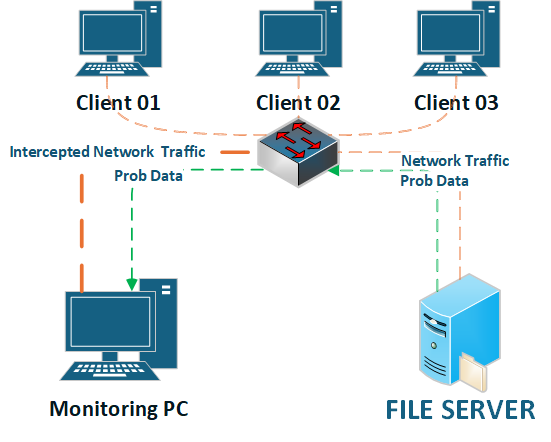}
    \caption{Implementation Scenario}
    \label{fig:3}
\end{figure*}

We started building a testbed where we execute samples safely. As depicted in Fig. \ref{fig:3}, we set up a client on a 64-bit Windows 10 Professional virtual machine with specs: 8 GB of memory, and 100 GB of SSD. The user does not have domain or local administrative rights. The user accesses the data folder through the mapped network drive. The other client PCs are for illustration purposes only and have not been used anywhere in a lab. 

We set up a  Windows 2016 virtual machine as a file server with specs: 8 GB of memory and 100 GB. The file server has been set up with a data folder of 200 files in 8 sub-folders. The root folder is shared with default read and write permissions, and the server is also acting as a domain controller and DNS. 

The monitoring PC is a 64-bit Ubuntu 24 Virtual Machine with 8 GB of memory and 50 GB of disk. All the virtual machines run on a 64-bit Ubuntu 24 host with specs: AMD RYZEN 5000 7 processor, 32 GB memory and 500 GB SSD and are connected to a virtual switch. The port where the Monitoring VM is connected has been configured in promiscuous mode to listen to packets between Client01 and the File Server. Behind the scenes, the host is connected to the WatchGuard Firewall, which continuously logs network access activities. 

\subsubsection{Ransomware Sample}\label{collection}

There are multiple publicly available sources of malware samples, including VirusShare, AnyRun, Malware Bazaar, etc. In order to download a sample, one must know the signature (usually hash values) of that malware. We searched multiple cybersecurity companies’reports for trending ransomware attacks and related indicators of compromise (IoC) in order to get malware hashes. We realised that the most recent ransomware strains' IoCs cannot be found anywhere other than in Virustotal. Unfortunately, downloading samples from VirusTotal is not possible unless someone has a premium subscription, which costs about 600 euros monthly. So, if any IoC cannot be found in other places, it is dropped.  We execute samples one by one in our testbed while capturing network traces between the client and file server for 10 minutes; a sample that does not encrypt the data folder is removed from the list (out of scope) because it might be human-operated ransomware, which requires manual operation via C2C. The final list of ransomware families covered in this work is presented in Table \ref{tab:6}.
\begin{table}
\caption{Ransomware Samples}
\label{tab:5} 
\begin{tabular}{|l|l|} \hline 

\textbf{Ransomware Group} & \textbf{Number of RoI} \\

\hline
ako & 8 \\
\hline
Coronavirus & 8 \\
\hline
netwalker & 41 \\
\hline
Avoslocker & 48 \\
\hline
Hive & 49 \\
\hline
ragnarok & 55 \\
\hline
Clop & 56 \\
\hline
Darkside & 58 \\
\hline
Wannacry & 62 \\
\hline
CryptoLocker & 78 \\
\hline
alpha & 79 \\
\hline
backcat & 81 \\
\hline
mountlocker & 81 \\
\hline
Conti & 87 \\
\hline
neffilim & 87 \\
\hline
nemity & 87 \\
\hline
DoppelPaymer & 88 \\
\hline
ransomexx & 92 \\
\hline
Revil/Sodinokibi & 110 \\
\hline
lockbit2.0 & 121 \\
\hline
maze & 121 \\
\hline
lockbit3.0 & 133 \\
\hline
\end{tabular}
\end{table}
\subsubsection{Benign Sample}
In a previous work, there are a few papers with details on the acquisition of benign samples. For example, Edwardo et al \cite{EduardoBerrueta2022} created benign samples by intercepting network traffic in a real production environment. Fernando et al \cite{Fernando2024} used VeraCrypt, AxCrypt, BitLocker and 7-Zip for benign. 

For our method, we used an extensive approach for generating benign samples. We try to mimic most of the possible normal user actions, which causes behaviour closer to that of ransomware, as explained below and presented in Table \ref{tab:6}. 
\begin{enumerate}
    \item Data Encryption.

    Practically, in a production environment where users access centrally hosted file servers, it is rare and unsafe to allow users to encrypt their data. The IT department would rather deploy such tools on the file server; that is why most commercial encryption tools are limited to local and extended disk partitions. However, there are a few tools that users can still use to encrypt data if required.
    
     \textbf{Hicrypt}; We searched and found that \textbf{Hicrypt} is one of the few tools that can encrypt files on a network share. This tool allows a user to create an empty folder within a network share and then encrypt it. All new files which will be added to that folder later will get encrypted automatically. So we empty our network drive and encrypt it using \textbf{Hicrypt }, then we import it into the same data folder while capturing network packets. 
   
    \textbf{Windows Encrypt Utility}; Windows Encryption Utility can encrypt both local and network files. Though rarely allowed in a production environment, we decided to capture the behaviour of this action so that if it happens for some reason, our detection model should remain accurate and precise.
    
   \textbf{WinRAR, Winzip,7Zip}; We realised that most file compression tools like WinRAR, WinZip and 7Zip have the option to encrypt compressed data. So we run these tools with an encryption option and then capture network traffic. The result of this operation is a compressed and encrypted file. We delete the compressed file before executing another tool.

    \item Bulk Data Import and Delete.

     Among the actions that users generally perform on a network share are importing a bunch of data from other storage. This action triggers behaviour closer to ransomware due to the fact that many new files get created, and a lot of data is written at the same time. To mimic this action, we used \textbf{Windows copy utility} as the most used tool in real life. We also used third-party tools \textbf{TeraCopy} and \textbf{UltraCopier} because some users prefer them over Windows utilities.   We empty the network drive, then paste it in the same folder being used for ransomware analysis and the other way around.
  
    \item Data compression.

    Sometimes, a user might need to transfer large amounts of files via email, ftp, for sharing with external partners. This can be achieved by compressing multiple files and folders into one. By doing that, a new file ( container ) is created, and files change extensions pretty much like what ransomware does when it changes the extension of all files after encryption and re-creates file contents. While there are many third-party tools for compressing files, we chose to use \textbf{WinRAR, WinZip, and 7-Zip} because they are the most used tools in the Windows environment. WinRAR and WinZip have output extensions that a user can choose from ( .zip and .rar). We run these tools while capturing network traffic.    
    \begin{table}[b]
\centering
\caption{Benign Samples}
\label{tab:6}
\begin{tabular}{|l|l|c|}
\hline
\textbf{Tool} & \textbf{Action} & \textbf{Number of RoIs} \\ \hline
\multirow{4}{*}{Windows Utility} 
               & Delete         & 11 \\ 
               & Copy-In        & 4  \\ 
               & Copy-Out       & 8  \\ 
               & Encrypt        & 50 \\ \hline
\multirow{3}{*}{WinRAR} 
               & Zip (.zip)     & 1  \\ 
               & Zip (.rar)     & 11 \\
               & Encrypt        & 4  \\ \hline
\multirow{2}{*}{WinZip} 
               & Zip (.zip)     & 8  \\
               & Encrypt        & 652 \\ \hline
\multirow{3}{*}{7-Zip} 
               & Zip (.zip)     & 8  \\ 
               & Zip (.7z)      & 4  \\ 
               & Encrypt        & 1  \\ \hline
TeraCopy       & Copy           & 43 \\ \hline
UltraCopier    & Copy           & 69 \\ \hline
HyCrypt        & Encrypt        & 1  \\ \hline
\end{tabular}
\end{table}

\end{enumerate}
\vspace{-3mm} 
\begin{table}[t]
\centering
\caption{Required SMB Headers}
\label{tab:2}
\setlength{\tabcolsep}{4pt} 
\renewcommand{\arraystretch}{1.2} 
\begin{tabularx}{\columnwidth}{|l|l|X|}
\hline
\textbf{Category} & \textbf{Field} & \textbf{Description} \\ \hline
Defaults & No & The index number of the packet \\ \hline
Defaults & Time & Time (seconds) elapsed since the capture started \\ \hline
Defaults & Source & Where the packet is coming from \\ \hline
Defaults & Destination & Where the packet is going \\ \hline
Defaults & Protocol & The type of protocol used for this communication \\ \hline
Defaults & Length & The length (in bytes) of the packet \\ \hline
Defaults & Info & The information carried in the packet (payload) \\ \hline
Additional & smb2-header-len & SMB header length \\ \hline
Additional & smb2-setinfo-size & The size of data contained in SMB Set Command \\ \hline
Additional & smb2-writecount & The size of data contained in SMB Write Command \\ \hline
\end{tabularx}
\end{table}

\subsection{Features Engineering} 
By default, Wireshark displays a limited set of fields for captured network traces, as shown in the Default Group Tab.\ref{tab:2}. To enhance the dataset for our experiment, it is necessary to include additional information beyond the default fields. To achieve this, we configure Wireshark to incorporate the extra fields listed in the Additional Group Tab.\ref{tab:2}.  Once the necessary fields are configured, we export the network traces in CSV format. This format is selected for its easy manipulation and efficient handling of the data during the subsequent stages of analysis and processing.

Since SMB commands are transmitted over TCP connections, duplicate packets can occasionally appear due to the connection-oriented nature of TCP. To address this, the module is configured to identify and remove duplicates. Some of the additional fields included in the preprocessing are specifically aimed at supporting this task, as the default fields alone may inadvertently lead to the removal of important packets. 

The final output of this pre-processing module is a clean CSV file, free from duplicates and containing all the necessary information required for further analysis in the subsequent steps.

\subsubsection{Region of Interest RoI}\label{RoI Technique}
Processing network traffic traces for machine learning applications presents significant challenges, as individual network packets typically contain limited information on their own. Therefore, a critical step in applying machine learning to network traffic is aggregating individual packets into more informative data samples. In this research, we introduce a novel method, referred to as the Region of Interest (RoI) technique, to address this issue. The RoI method aims to aggregate related packets into cohesive data samples, ensuring that the aggregated information captures the necessary context for downstream analysis and feature extraction.
\begin{algorithm}
\caption{Extracting $RoI$ Techniques}\label{alg:3}
\begin{algorithmic}[1]
\Require $ \mathcal{P}  == Network \quad Packets \quad flows$
\State Let $t_i , t_j  Represent\quad two \quad consecutive \quad packet p_i and p_j $
\State Let $p260 \quad Represent\quad a \quad packet \quad with \quad length \quad == 260 \quad bytes $
\State Let $p \neq p260\in   \mathcal{P}$, where\quad d \quad is\quad a \quad network \quad packet \quad other \quad than \quad $p260$

 \State  \hspace{4em} $ ROI_i \gets \forall p \in \{t_i,j_j\} $

\end{algorithmic}
\end{algorithm}
Server Message Block (SMB) messages typically consist of data operation commands for session setup and termination, file opening, reading, writing, status checks, and their respective responses \cite{opcode}. Our research observed that the SMB client frequently navigates through directories, enumerating contents in a back-and-forth manner. This behaviour is facilitated by a specific packet, which remains consistent in size (260 bytes) regardless of the environment or the directory paths being accessed.  It serves as an indicator for an SMB client process, signalling it whether all items in the current directory have been enumerated and whether the client needs to traverse back or forward into other sub-directories. We refer to this packet as $p260$ and is extracted using the algorithm in \ref{alg:3}.

The procedure outlined in Algorithm \ref{alg:3} requires a set of network packet traces collected from the time the first $p260$ appears to the subsequent $p260$, excluding the latter. Consequently, we define the Region of Interest (RoI) as the set of activities between two consecutive $p260$ packets.

The encryption process takes plain text as input into the cypher algorithm in order to produce a ciphertext. That means, the original file contents are replaced with cypher text. Ransomware goes the extra mile by changing the extension of the file. We discovered that the mission is completed using special SMB commands referred to as Commands of Interest ($CoI$) as explained below.
  \begin{enumerate}
    \item 
    At some point in time, a client wants to read a file that does not exist yet; So it will create it, and the server shall reply confirming that a new file has been created. This response/backward packet has been found with a static length of (410) regardless of the environment or file path. This  {$CoI$}is referred to as ${rp410}$.
    \item 
    At some point in time, the client will read any file in the current directory; this forward packet has been found with a static length of (171) regardless of environment or file path. This {$CoI$} is referred to as ${rq171}$
    \item 
    At some point in time,, a client will write data to files in the current directory and the server shall reply with the number of bytes written to that file. This backward/response packet has been found with a static length (138) regardless of the environment or file path. This {$CoI$} is referred to as ${rp138}$.
     \item 
    At some point in time, the client will attempt to open a file/folder, and the server responds with ${ rp378}$. 
   \item 
   At some point in time, the client wants to set new information on files, such as renaming. This forward packet has been found with a static length of (124) regardless of the environment or file path. This {$CoI$} is referred to as ${rq124}$.
    \item 
    At some point in time, the client will close the open file/folder. This forward packet has been found with a static length of ${ rq146}$.
  \end{enumerate}
\vspace{-3mm} 
\begin{table}[b]
\centering
\caption{Extracted Basic Features}
\label{tab:3}
\setlength{\tabcolsep}{4pt} 
\renewcommand{\arraystretch}{1.2} 
\begin{tabularx}{\columnwidth}{|c|X|c|}
\hline
\textbf{Feature} & \textbf{Description} & \textbf{Related $COI$} \\ \hline
Source & The origin of a packet (IP address) & n/a \\ \hline
Destination & The destination of a packet (IP address) & n/a \\ \hline
Duration & Activity duration & n/a \\ \hline
Read & Number of files read & 171 \\ \hline
Open & Number of files and directories opened & 378 \\ \hline
Modify & Number of newly created files & 410 \\ \hline
Write & Number of files with content modified & 138 \\ \hline
Set & Number of files with modified attributes & 124 \\ \hline
Close & Number of files and folders closed & 146 \\ \hline
Label & Ransomware family or Benign Application Name & n/a \\ \hline
\end{tabularx}
\end{table}
\begin{algorithm}
\caption{Extended Features Computation Techniques}
\label{alg:4}
\begin{algorithmic}[1] 
\Require $ {RoI} $
\State Let $CTW = \emptyset$;  Stands for  Create  Then  Write  packets as results of  $p410$ 
\State Let $RTW == \emptyset$; Stands for Read Then Writes packet as results of $p171$
\State Let $c\in RoI=\{ 410 , 171\}$; a set of centroids  packets 
\State Let $ rp138(i)i_\in\{1,...,n\}$; any  Write  response  packet  within  $RoI$
\State For each $rp138_i:$
\If{$\{(rp138_i - c_0)== Minimum\} $}  
    \State $CTW   \gets rp138_i[writecount]$
\EndIf
\If{$\{(rp138_i - ci_1)== Minimum\}  $}  
    \State $RTW   \gets C138_i[writecount]$ 
\EndIf 
\State $Let x \in  RTW $
\State $Let y \in  CTW $
\State $Let \mu, mean \quad of \quad x \quad or \quad y $
\State X ={$\sum_{}^{}(x_i)$}
\State $TotX= {\sum_{0}^{i}(x[writecount])}$
\State $\mu_x$ ={$\frac {TotX} {X}$}

\State Y ={$\sum_{}^{}(y_i)$}
\State $TotY= {\sum_{0}^{i}(y[writecount])}$
\State $\mu_y$ ={$\frac {TotY} {Y}$}
\State sdvX={$\sqrt{\frac {\{\sum_{}^{}(x_i-\mu_x)^2 \hspace{1em}} {X}} $}
\State sdvY={$\sqrt{\frac {\{\sum_{}^{}(y_i-\mu_y)^2 \hspace{1em}} {Y}} $}
\State $ MRW_x== x[writecount]_i \in X , most frequent $
\State $ MRW_y== y[writecount]_i \in Y , most frequent $
\end{algorithmic}
\end{algorithm}

Based on \textbf{$CoI$} as explained above, we compute basic features ($BF$) using algorithm \ref{alg:5}. For each  \textbf{$RoI$}, we count the frequency of each \textbf{$CoI$}.

\begin{algorithm}
\caption{Basic Features Computation Procedure }
\label{alg:5}

\begin{algorithmic}[1]
\Require $  RoI $
\State Let $ l; a \quad a packet \quad length \quad header \quad size $
\State Let $ P; \hspace{1em}  P(l)\notin \{410,171,138,124,146\}  ; a \quad a packet $
\For{each $CoI \in\{410,171,138,124,146\} $}    
\State \hspace{1em}  $BF=\forall P ,\hspace{1em} P(l)==CoI $
    
\EndFor
 \end{algorithmic}  
\end{algorithm}


Normally, you can save new data by creating a new file, writing on it, then saving, or by opening an existing file, writing additional data, then saving. The relevant \textbf{$CoI$} as earlier discussed are \textbf{$rq171$},\textbf{$rp410$} and \textbf{$rp138$}.
\textbf{$rp138$} tells us that new data has been successfully saved either following the modification on an existing file  \textbf{$rq171$} or by creating a new one  \textbf{$rp410$}. 

As depicted in algorithm \ref{alg:4}, for each Write response packet, we find whether it has been triggered by reading an existing file ($rq171$) or by creating a new file ($rp410$). The latest trigger to occur before that write response is considered (the closest command to the write response). Next, from the extracted $CTW$ and $RTW$, we compute total, mean, and Standard Deviation. The features as results of this algorithm are explained in Table \ref{tab:4}

\begin{table}[t]
\centering
\caption{Computed (Extended) Features}
\label{tab:4}
\setlength{\tabcolsep}{4pt} 
\renewcommand{\arraystretch}{1.2} 
\begin{tabularx}{\columnwidth}{|l|X|l|}
\hline
\textbf{Feature} & \textbf{Description} & \textbf{Related $COI$} \\ \hline
rtw\_cnt   & Number of write responses triggered by read & C171 \\ \hline
rtw\_tot   & Total written data triggered by read & C171 \\ \hline
rtw\_mean  & Mean of written data triggered by read & C171 \\ \hline
rtw\_stdv  & Standard deviation of written data triggered by read & C171 \\ \hline
rtw\_mrw   & Most repeating written data triggered by read & C171 \\ \hline
ctw\_cnt   & Number of write responses triggered by create & C410 \\ \hline
ctw\_tot   & Total written data triggered by create & C410 \\ \hline
ctw\_mean  & Mean of written data triggered by create & C410 \\ \hline
ctw\_stdv  & Standard deviation of written data triggered by create & C410 \\ \hline
ctw\_mrw   & Most repeating written data triggered by create & C410 \\ \hline
\end{tabularx}
\end{table}
\subsubsection{Features Standardization}

The cumulative values obtained as results of algorithm \ref{alg:4} vary from one production environment to another, as the structure and size of the shared network directory are different and have different infrastructure setups. In order to generalise our method for different environments, we introduced a script to be installed on the file server for constantly reporting statistics of files and directories available in the shared network drive (Probing). We use the probe information; the number of files  ($n$) of the shared drive at the time of capture to convert basic features ($BF$) into rates based on $100$ files as per equation \ref{eq:1}. The number of features (fields) computed and what each stands for is presented in Table \ref{tab:3}. 
\begin{equation} \label{eq:1}
 \mathcal{SF} =\frac {\mathcal{BF}*100} {n } \hspace{1em} 
\end{equation}
\subsection{Train  Machine Learning Model} \label{ML}
In this sub-section, we use the dataset that we created as per previous sub-sections to create a learning model that detects the ransomware based on historical sample data. The original dataset from the testbed is imbalanced with $844$ benign and $1600$ ransomware data samples.  
\begin{itemize}
\item \textbf{Data Sanitation}. We removed data samples corresponding to $IoC$. We do this because these data have well-known and distinct characteristics for all ransomware families that cannot be found anywhere with benign samples. These data are collected in a separate repository of  $IoC$ as previously explained to serve for $IoC$ detection phases. This results in reduced ransomware data samples up to $1262$.  
\item \textbf{Data Balancing}
As it is always a good idea to use a balanced dataset when training an ML model, we used a Python function to select random $844$ from $1262$ ransomware samples. This task was done while the ransomware samples still had labels (Ransomware Family Name) to make sure that all families are represented. Then we labelled instances of the resulting sub-dataset as ransomware.

\item \textbf{Training Vs Testing}.
Supervised Machine Learning, required labelled sample data for training and unlabeled ones for testing. This can be achieved by splitting the whole dataset into training and testing sub-datasets or by using K-fold cross-validation \cite{K-FOLD}. Splitting a dataset is most of the time achieved by random sampling. The problem with the method is that a single run may select irrelevant data that influences biased learning. K-fold cross-validation techniques, on the other hand, split the dataset into multiple sample groups (K), resulting in a less biased learning model. So in this research, we used 10-fold cross-validation with different.

\item \textbf{Evaluation}.
We tested the dataset with different traditional Machine Learning techniques and evaluated the performance of each. Unlike in other domains where ML is applicable, high False negative rates expose the business to higher risk than False Positive rates. In other words, a benign application detected as malicious would only cause time waste, leaving company data intact. But a malicious application passed unseen will cause data/system damage, and financial loss and cost more time for recovery  \cite{Maniriho2022}. While most of the previous work evaluated ML models based on accuracy and False Positive rates as per equation  \ref{eq:2} and \ref{eq:3}, respectively, in this research, we extend the evaluation to include false negative rate metrics as per equation \ref{eq:4}.
\end{itemize} 
\begin{equation}\label{eq:2}
{Accuracy= \frac{(fp+tp)}{(tn+fp+tp+fn)} \hspace{1em} }
\end{equation}

\begin{equation}\label{eq:3}
{FPR= \frac{(fp)}{(fp+tn)} \hspace{1em} }
\end{equation}

\begin{equation}\label{eq:4}
{FNR= \frac{(fn)}{(fn+tp)} \hspace{1em} }
\end{equation}
where;
\\
$fp$ == number of  Normal classified as Ransomware \\
$fn$== number of  Ransomware classified  as  Normal \\
$tp$== number of Ransomware  classified as  ransomware \\
$tn$== number of  Normal classified as Normal 

\section{Results and Discussion}
In this section, we analyse the experimental results and their contribution to advances in ransomware detection.

\textbf{Robustness}.
We tested the dataset with different machine Learning Models. As the results show in the Table \ref{tab:7}, Random Committee classification exhibits the best performance with  99.6\% accuracy,  0\% False Negative Rate (FNR) and 0. 004\% False Positive Rate (FPR). This result represents the best performance achieved in our experiments, demonstrating both robustness and high efficacy.
\begin{table}[h!]
\caption{Performance Metrics of Various Classifiers}
\label{tab:7}
\centering
\begin{tabular}{|l|c|c|c|}
\hline
\textbf{Classification} & \textbf{Accuracy} & \textbf{FPR} & \textbf{FNR} \\
\hline
RandomCommitee & 99.645 & 0.004 & 0.001 \\
\hline
RF & 99.585 & 0.004 & 0.001  \\
\hline
Rules.Part & 98.874 & 0.011 & 0.010 \\

\hline

lazy.ibk & 98.519 & 0.015 & 0.015 \\
\hline
BayesNet & 98.045 & 0.020 &0.016  \\
\hline
Logistics & 97.275 & 0.027 & 0.022 \\

\hline
\end{tabular}
\end{table}

For comparison purposes, we also create models for only 1 (Lockbit2.0) and two (Lockbit2.0 \& Maze) ransomware families, as has been done in \cite{Almashhadani2022}. In both scenarios, the accuracy achieved is 100 \% accuracy. However, we caution that such models are not robust due to the limited scope of ransomware families considered. Given the large and evolving number of ransomware families, we do not recommend this approach for broader applications.

\textbf{Early Detection }. Early detection refers to a kind of model in which dynamic features are captured before the executed payload completes all tasks. The objective is to detect intrusion before it causes the most damage \cite{Cen2024}. Different scholarly articles, such as \cite{Deng2023}, proposed methods exclusive to ransomware early detection on a host-based basis. Berrueta et al.\cite{Morato2018} proposed a method using a threshold for early detection of ransomware based on SMB network traffic.

\begin{figure*}[b!]
    \centering
    
    \begin{subfigure}{0.48\textwidth}
        \centering
        \includegraphics[width=\linewidth]{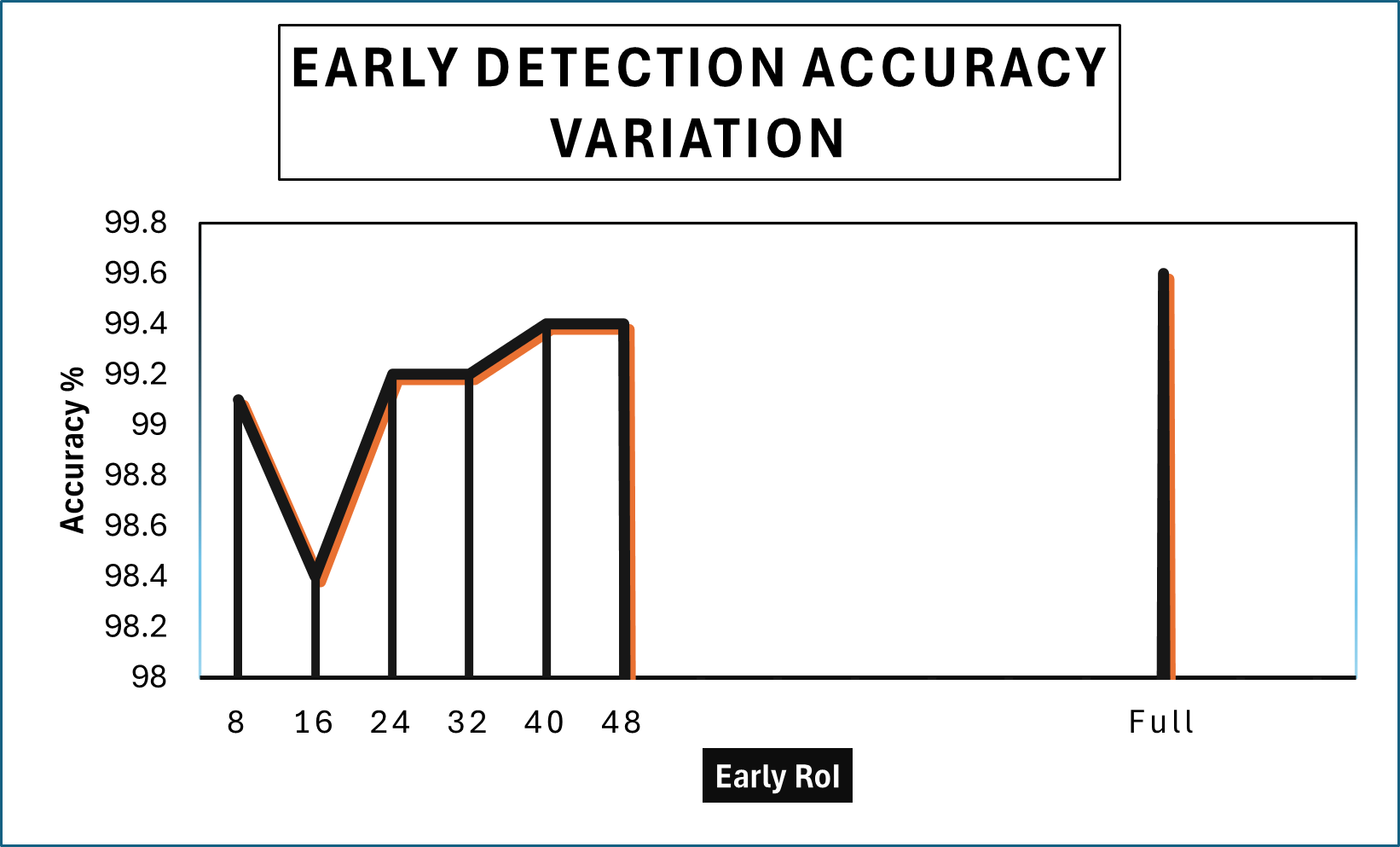}
        \caption{Early Detection Accuracy}
        \label{fig:4}
    \end{subfigure}
    \hfill
    \begin{subfigure}{0.48\textwidth}
        \centering
        \includegraphics[width=\linewidth]{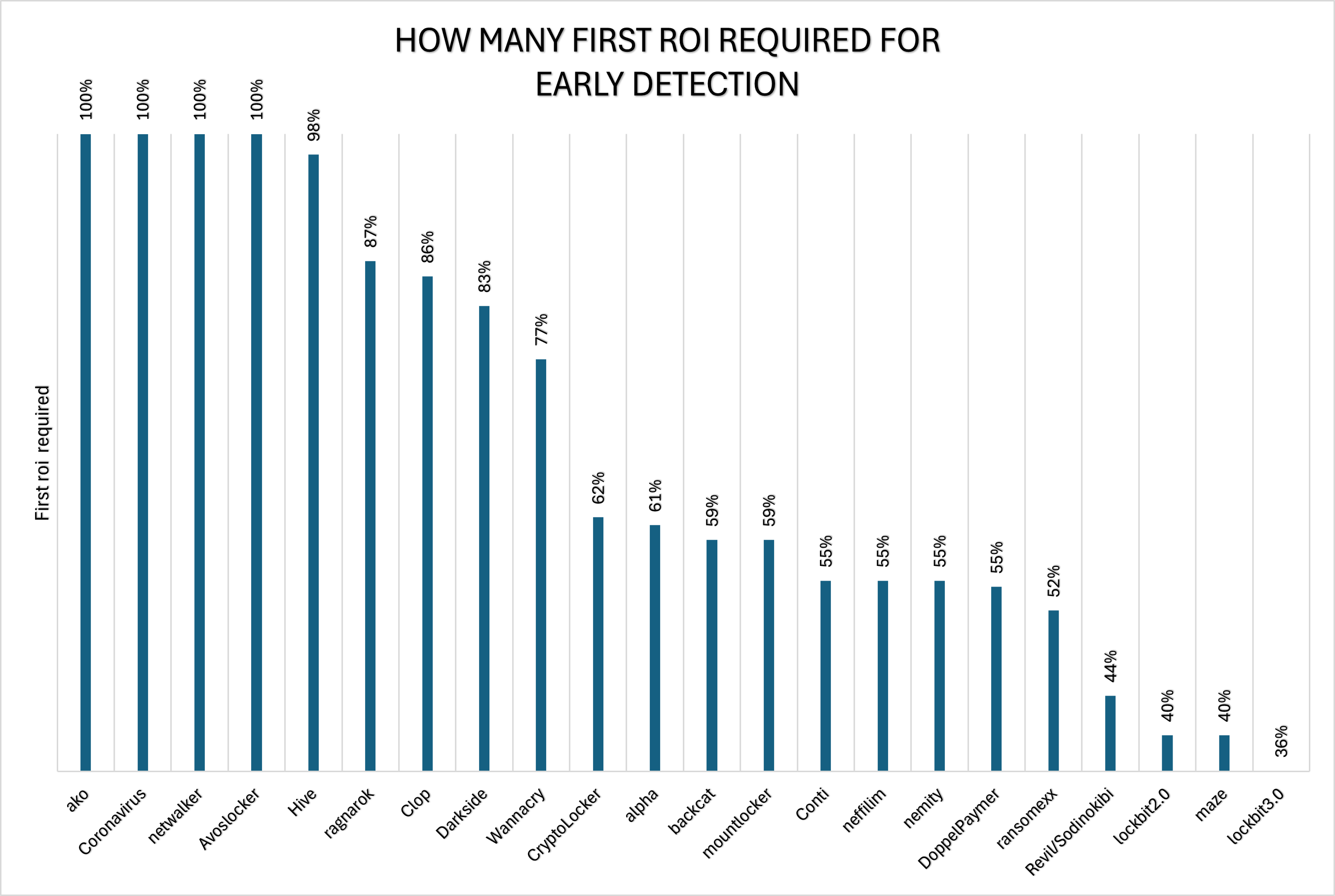}
        \caption{Early Detectability}
        \label{fig:5}
    \end{subfigure}
    
    \caption{Comparison of early detection performance metrics}
    \label{fig:early_metrics}
\end{figure*}

In our dataset, the smallest ransomware data sample size is 8 for Ako and Coronavirus. So, for each ransomware family, we first created a sliding window of 8 $RoI$. That means that the first test scenario has $176(8x22)$, the second has $352 (16x22)$, etc. For each test scenario, we randomly selected an equivalent benign sample from a benign dataset and then classified the data in the same way as for full detection. The window size that produces high accuracy is considered the best for early detection. As shown in Fig. \ref{fig:4}, the first 48 $RoI$ produced 99.44 \% accuracy, just 0.24 \% away from full detection accuracy.

Since different ransomware families have different numbers of $RoI$, it is obvious that the degree to which each of the families can be detected at an early stage is different. For example, Fig. \ref{fig:5} shows that Lockbit can be fully detected with 36 \% of its execution. Only Ako, coronavirus, netwalker and avoslocker require completion execution as they produced quite a small number of $RoI$s. 

\textbf{Dataset and $IoC$ Repository}.
As part of the experimentation results, we created a $RoI$ dataset from 22 ransomware families and 15 benign applications. Benign and Ransomware samples have all been collected from the same environment (testbed). Benign samples cover an extensive number of benign applications to mimic actions that a human can perform, nearly similar to what ransomware does. We found and extracted a new network-based $IoC$ that we stored in a repository for use with signature or rule-based detection techniques.

\section{Conclusion  \& Future Work}
In this paper, we present a new framework for detecting crypto ransomware through the application of Region of Interest (RoI) techniques to SMB network traffic. The framework operates in three distinct phases: the first two phases function in a signature-based mode, inspecting individual and aggregated packets against known Indicators of Compromise (IoCs), while the third phase employs the Random Committee Learning Model, achieving an impressive accuracy of 99.64\%, with no ransomware left undetected. Furthermore, we evaluated the machine learning model for early detection and achieved a notable accuracy of 99.44\%.

Additionally, our study has led to the discovery of new IoCs, which can be utilised in signature- and rule-based detection systems. This contribution holds substantial significance both academically and industrially. The experiments were conducted on SMB3 for both encrypted and unencrypted network traffic, with the presented IoCs corresponding to unencrypted traffic. However, each code has a counterpart for encrypted traffic. In future work, we aim to expand the analysis to include additional ransomware families and continually update our datasets.

\bibliographystyle{cas-model2-names}
 \bibliography{library}

\end{document}